\documentclass[aps,prb,preprint,superscriptaddress]{revtex4}
\usepackage[english]{babel}
\usepackage{epsfig}

\begin{document}

\title{Orbital occupation, atomic moments and magnetic ordering at interfaces
of manganite thin films}
\author{C. Aruta}
\email{aruta@na.infn.it} \affiliation{CNR-INFM Coherentia,
Dipartimento di Scienze Fisiche, Universit$\grave{a}$ di Napoli
``Federico II'', Complesso di Monte S. Angelo, Via Cinthia,
I-80126 Napoli, Italy}
\author{G. Ghiringhelli}
\affiliation{CNR-INFM Coherentia and Soft, Dipartimento di Fisica,
Politecnico di Milano, piazza Leonardo da Vinci 32, I-20133
Milano, Italy}
\author{V. Bisogni}
\affiliation{European Synchrotron Radiation Facility, Bo\^{\i}te
Postale 220, F-38043 Grenoble, France}
\author{L. Braicovich}
\affiliation{CNR-INFM Coherentia and Soft, Dipartimento di Fisica,
Politecnico di Milano, piazza Leonardo da Vinci 32, I-20133
Milano, Italy}
\author{N.B. Brookes}
\affiliation{European Synchrotron Radiation Facility, Bo\^{\i}te
Postale 220, F-38043 Grenoble, France}
\author{A. Tebano}
 \affiliation{CNR-INFM Coherentia and Dipartimento di Ingegneria
Meccanica, Universit\`{a} di Roma Tor Vergata, Via del Politecnico
1, I-00133 Roma, Italy}
\author{G. Balestrino}
\affiliation{CNR-INFM Coherentia and Dipartimento di Ingegneria
Meccanica, Universit\`{a} di Roma Tor Vergata, Via del Politecnico
1, I-00133 Roma, Italy}

\date{\today}

\begin{abstract}
We have performed x-ray linear and circular magnetic dichroism
experiments at the Mn $L_{2,3}$-edge of the
La$_{0.7}$Sr$_{0.3}$MnO$_{3}$ ultra thin films. Our measurements
show that the antiferromagnetic (AF) insulating phase is
stabilized by the interfacial rearrangement of the Mn $3d$
orbitals, despite the relevant magnetostriction anisotropic effect
on the double-exchange ferromagnetic (FM) metallic phase. As a
consequence, the Mn atomic magnetic moment orientation and how it
reacts to strain differ in the FM and AF phases. In some cases a
FM insulating (FMI) phase adds to the AF and FM. Its peculiar
magnetic properties include in-plane magnetic anisotropy and
partial release of the orbital moment quenching. Nevertheless the
FMI phase appears little coupled to the other ones.


\end{abstract}
\maketitle

\section{Introduction}
Interfaces obtained by assembling insulating, non-magnetic
perovskite oxides can show unexpected properties such as high
conductivity and ferromagnetism \cite{1,2}. Oxygen vacancies
\cite{3}, epitaxial strain \cite{4,5}, the so-called
``polarization catastrophe'' from interface-generated dipoles
\cite{6} and electronic reconstruction at the interface \cite{7,8}
are all at play in perovskite oxides and their individual roles
are still far from being understood. An important case is the
interface of manganites thin films with other different oxides,
providing tunnel junctions for a number of manganite-based
devices, such as spin valve or spin injectors. In this context,
several authors have investigated the properties of ultra-thin
manganite films on various substrates and it has been found at the
double exchange (DE) magneto-transport properties are strongly
depressed below a critical thickness \cite{9}. To explain such a
behavior, nanoscale inhomogeneities with coexisting clusters of
different stable phases have been extensively investigated, but it
is still uncertain if and how this acts on the suppression of the
magneto-transport properties of ultrathin films \cite{18}.
Chemical composition, strain and oxygen stoichiometry are
considered the main parameters influencing the disorder-driven
phase separation \cite{Dagotto}. In thin films of
La$_{0.7}$Sr$_{0.3}$MnO$_{3}$ (LSMO) the existence of intrinsic
inhomogeneities has been explained in terms of structural
macroscopic distortions induced by the strain with the substrate
which favors preferential orbital occupation of the $e_{g}$ Mn
orbitals \cite{ArutaPRB, TebanoPRB}. While in the case of LSMO
films grown on LaAlO$_{3}$ substrate the suppression of the DE
magneto-transport properties is a ``bulk'' effect caused by strong
in-plane compressive epitaxial strain, in the case of the LSMO
films grown on SrTiO$_3$ (weak tensile strain) and NdGaO$_{3}$
(almost unstrained) the same phenomenon is a pure
``interface/surface'' effect \cite{TebanoPRL}. Broken symmetry at
the interfaces to the substrate and to the vacuum drives the
orbital reorganization in ultrathin LSMO films, thus favoring the
occupation of the $e_{g}(3z^{2}-r^{2})$ versus the $e_{g}
(x^{2}-y^{2})$ orbitals among the otherwise energy degenerate Mn
$3d$ states at the Mn$^{3+}$ sites, as shown in Fig.~1. In
addition, it has been recently reported \cite{Herger} that
structure and stoichiometry gradually change at the interface of
ultrathin LSMO films on STO, with a resulting elongation of the
interfacial out-of-plane lattice constant. Such structural
modification is similar to the cooperative Jahn-Teller-like
distortion induced by the in-plane compressive strain, which in
turn favors the stabilization of the $e_{g}(3z^{2}-r^{2})$
orbitals. As a result, the disproportion in $e_{g}$ orbital
occupation induces a coupling between neighboring Mn cations that
is ferromagnetic (FM) along the $c$-axis (perpendicular to the
surface) and antiferromagnetic (AF) in the $ab$ plane \cite{14},
eventually resulting in the stabilization of the C-type AF phase
at low temperature \cite{15}. However, such magnetic phase was not
directly experimentally observed.
\begin{figure}
\includegraphics[width=8 cm]{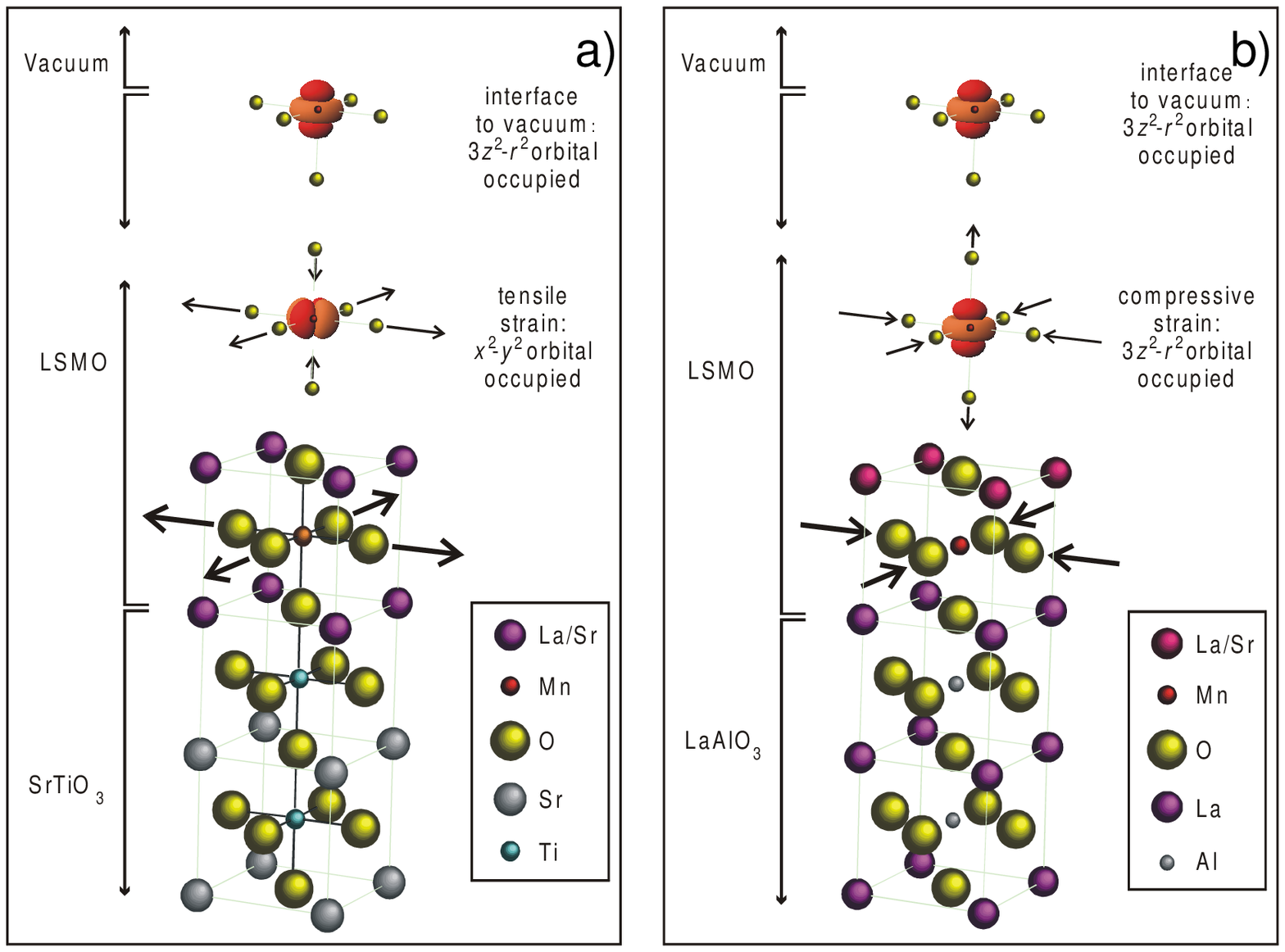}
\caption{(Color online) Schematics of the influence of strain onto
the orbital occupation in the bulk and at the interface of LSMO
films. The bulk Mn $3d$ orbital occupation is $x^{2}-y^{2}$ and
$3z^{2}-r^{2}$ in case of SrTiO$_3$ (a) and LaAlO$_{3}$ (b)
substrate respectively. The preferential orbital occupation is
$3z^{2}-r^{2}$ at both interfaces.}
\end{figure}

While the ``bulk'' magnetic properties have been largely
investigated in manganite films as a function of the
strain\cite{Song, Wu, Dvorak, Nath}, the microscopic origin of the
magnetic properties at the films surface and at the interface
between film and substrate have not yet been completely clarified.
 In this respect experimental investigations by surface sensitive
x-ray magnetic scattering of layered manganite single crystals
\cite{16} and of perovskite manganite thin films \cite{17} have
shown that the average in-plane FM ordering of the surface is
significantly suppressed over a length scale of about 4 unit cells
(u.c.) from the surface, even at low temperature. However, those
measurements alone cannot discriminate between a homogeneous
suppression of the magnetization over the surface and the
coexistence of FM and AF phases. Thus, to detect both AF and FM
phases, we have chosen two magnetically complementary techniques
such as x-ray magnetic linear dichroism (XMLD) and x-ray magnetic
circular dichroism (XMCD) in soft x-ray absorption spectroscopy
(XAS) by synchrotron radiation. Such techniques were already
successfully employed to observe the directional coupling by
exchange bias between the spins in the AF regions and those in the
adjacent FM regions in different magnetic systems\cite{Nolting}.
Therefore, XMLD and XMCD are the ideal techniques to study the
arrangement of spins, together with the orbital occupancy, at
interfaces in manganite films where AF and FM phases coexist on a
nanometric scale. While strain induced selective orbital occupancy
has been recently reported in refs \cite{Abad, Infante}, the
orbital reconstruction at the interface has been questioned by
Huijben \textit{et al}.\cite{Huijben}, on the basis of apparently
very different linear dichroism experimental results. Actually the
XLD of ref \cite{Huijben} is not totally incompatible with that of
ref. \cite{TebanoPRL} if one takes into account the fact that the
former were measured at low temperature, where the magnetic
contribution to XLD is strong, whereas the latter were taken above
the Curie temperature, where the only contribution to XLD comes
from the preferential orbital occupation. Moreover the
interpretation given by Huijben \textit{et al}. of the XLD is
surprisingly opposite to that of numerous papers with experimental
and theoretical contents \cite{ArutaPRB, TebanoPRB, TebanoPRL,
theoXLD}. Interestingly, Huijben \textit{et al}.\cite{Huijben}
have also reported that the spin to orbital-ordered coupled
insulator phase develop at the interface. Therefore, the
microscopic origin of the magneto-transport properties at the
interface of LSMO films is still under debate. Here we report the
experimental evidence of the spin-orbit-lattice coupling at the
interface of LSMO films. To achieve this result, we have compared
LSMO films grown on SrTiO$_3$ (100) (STO), NdGaO$_{3}$ (110) (NGO)
or LaAlO$_{3}$ (100) (LAO), and having different thicknesses so to
have different strain conditions and to cross the metal-insulator
transition at different temperatures. The high surface sensitivity
of both XMLD and XMCD allowed to obtain information on the very
thin LSMO layer at the interface with the substrate.
\begin{table}
\caption{Out-of-plane ($\epsilon_{zz}$) and in-plane
($\epsilon_{xx}$) strain values and metal-insulator transition
temperatures ($T_{MI}$) for LSMO films grown on the different
substrates with different thickness. The samples are listed by
decreasing transition temperature.}
\begin{ruledtabular}
\begin{tabular}{c c c c c}
Substrate & Thickness (u.c.)& T$_{MI}$(K)& $\epsilon_{zz}$($\%$) & $\epsilon_{xx}$($\%$)\\
LAO  & 100 & 360 & 1.30 & -1.70  \\
STO  & 50 & 360 & -1.10 & 0.90  \\
STO  & 10 & 275 & -1.10 & 0.90 \\
NGO  & 9 & 200 & 0.20 & -0.20  \\
LAO & 30 & - & 3.60 & -2.20 \\
\end{tabular}
\end{ruledtabular}
\end{table}
\section{Experimental}
LSMO films were grown by pulsed laser deposition with in situ
reflection high energy electron diffraction (RHEED). Film
thickness was controlled at the level of a single unit cell by the
intensity oscillations of the RHEED specular spot. Additional
details on the growth technique are given in ref.
\cite{TebanoEPJ}. The crystallographic and transport properties of
the investigated samples, obtained by x-ray diffraction and
electrical measurements\cite{ArutaPRB, TebanoPRB,TebanoPRL},
are reported in Table I. The out-of-plane ($\epsilon_{zz}$) and
in-plane ($\epsilon_{xx}$) strains are defined as the percentage
variation of the out-of-plane and in-plane lattice parameters of
films relative to the bulk LSMO values. From Table I it can be noticed that, in the case of the LAO substrate, the 30 u.c thick film is fully strained (in-plane compressive) while strain is
partially relaxed in the 100 u.c. thick film. Films on STO substrates are fully strained (in-plane tensile) regardless of film thickness. Finally, because of the good lattice
match with the substrate, thin films on NGO result to have lattice parameters only slightly distorted relative to the bulk. The metal-insulator transition temperatures
($T_{MI}$) reported in Table I demonstrate that the suppression of
the magnetotransport properties is strain dependent in case of
LAO, but is an interface
effect in case of STO and NGO\cite{TebanoPRL}.\\
Linear and circular dichroism measurements were carried out at the
ID08 beam line of the European Synchrotron Radiation Facility
(ESRF) by tuning the synchrotron radiation at the Mn L-edge. The
dominant photon-excited transitions are $2p \rightarrow 3d$, as
detected by total electron yield. Spin-orbit interaction splits
the L-edge absorption spectra into the $L_{3}$ and $L_{2}$ edges
with opposite spin-orbit coupling ($l+s$ and $l-s$, respectively).
A reversible and tunable (up to 1 T) external magnetic field can
be used to modify the dichroic response of the sample.\\
Circular dichroism is the difference in the absorption of photons
with right-handed or left-handed circular polarization and linear
dichroism (XLD) is the difference in the XAS measurement when the
electric vector of the incident photons is rotated by
$90^{\circ}$, by using synchrotron radiation with horizontal (H)
and vertical (V) polarizations. In all the XAS measurements a
constant background was fitted to the pre-edge region of the
$L_{3}$ edge and subtracted from the spectra, which are then
normalized to the edge jump set to unity above the $L_{2}$ edge.

\begin{figure}
\includegraphics[width=8 cm]{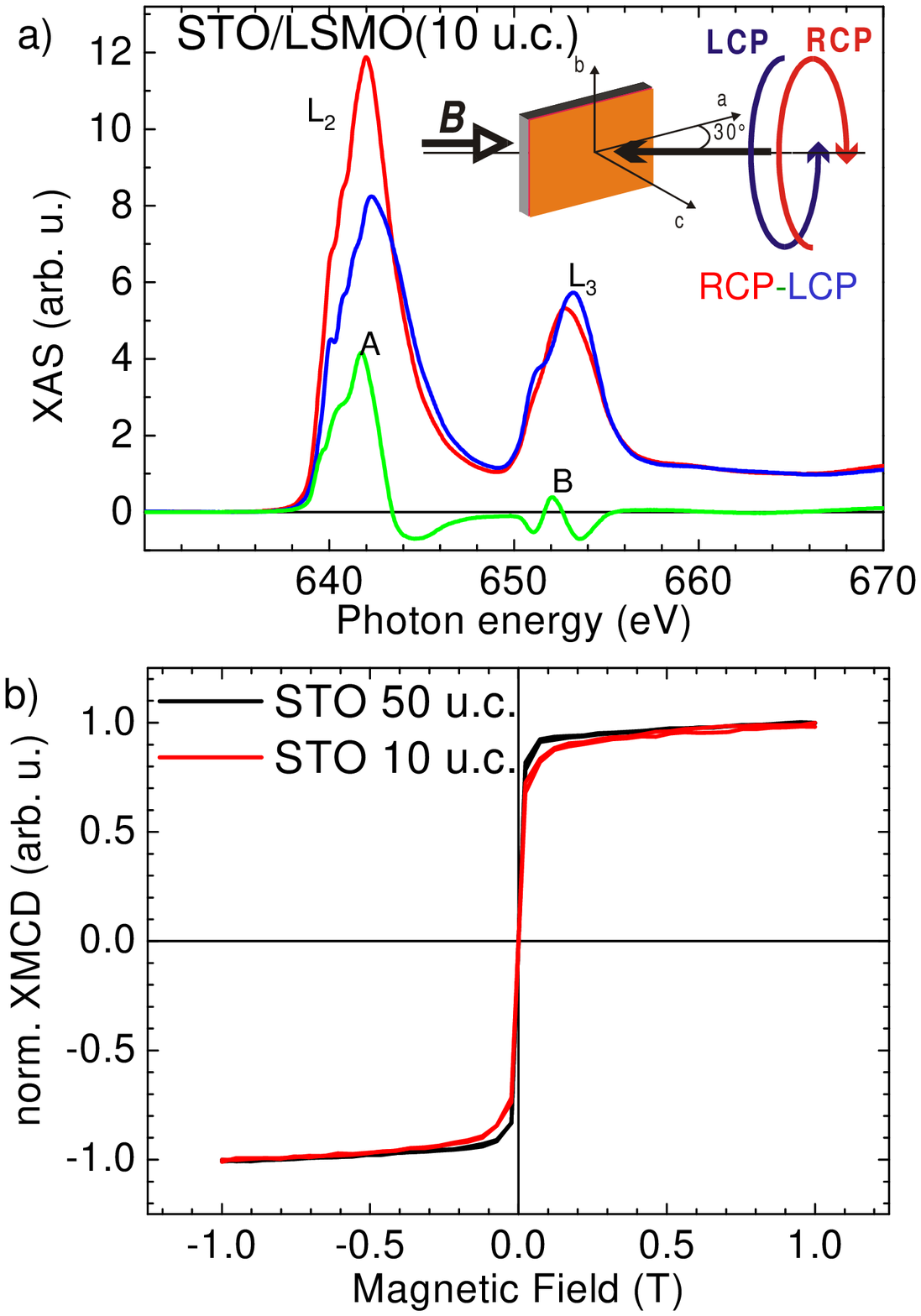}
\caption{(Color online)(a) Typical XAS and XMCD  results for 10
u.c. thick film on STO with an applied magnetic field of $B=1T$.
In the inset the experimental configuration is shown. The XMCD
results are reported as a difference of the XAS measurements with
Right (RCP) and Left (LCP) polarizations and without any further
normalization. (b) Hysteresis loops curves between $B = -1T$ and
$B = 1T$, for 50 u.c. and 10 u.c. thick films on STO. The curves
are normalized to unity for a better comparison of the coercitive
fields. All measurements in (a) and (b) were performed at
temperature of T=10K.}
\end{figure}

\begin{figure}
\includegraphics[width=8 cm]{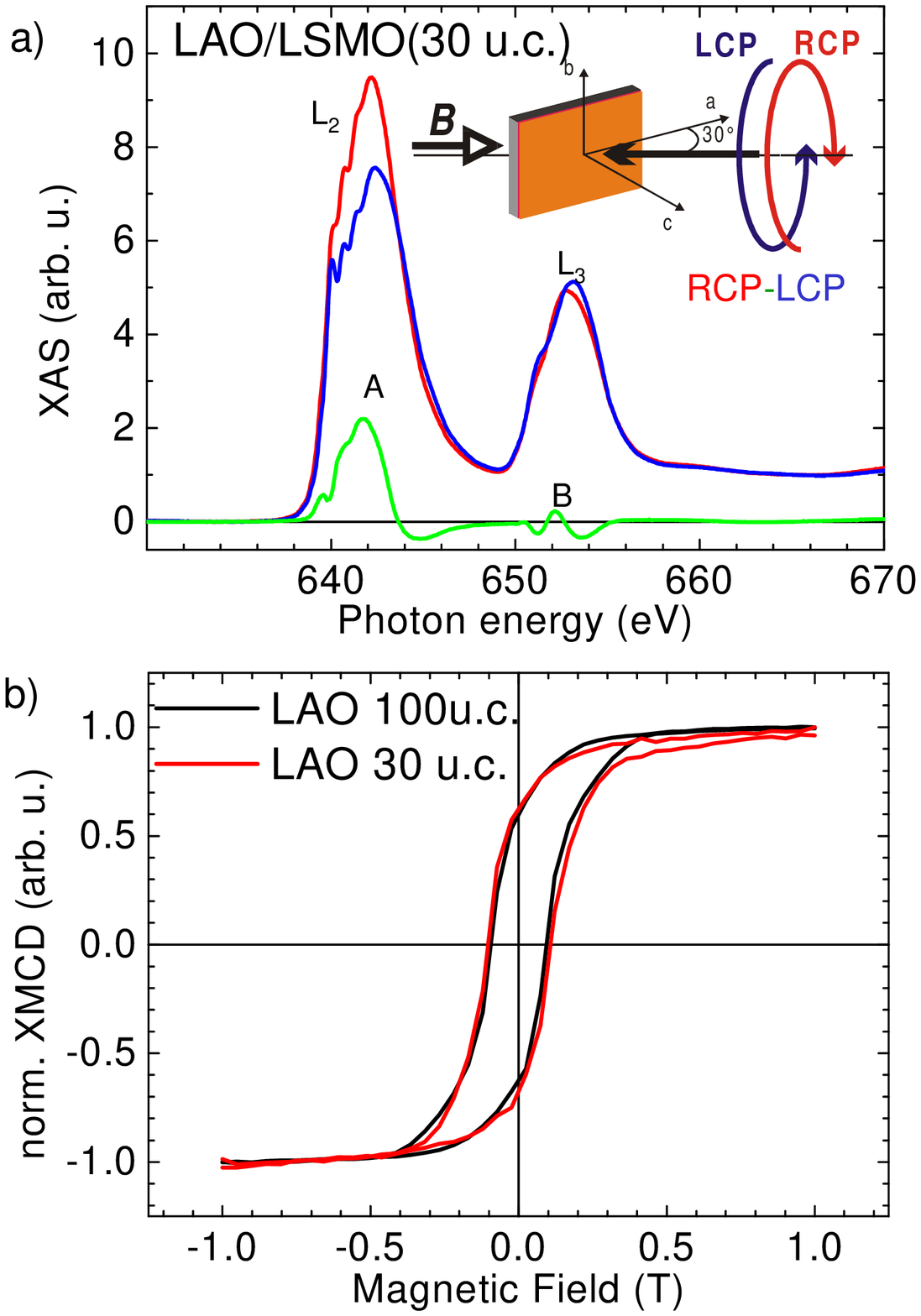}
\caption{(Color online)(a) Typical XAS and XMCD  results for 30
u.c. thick film on LAO with an applied magnetic field of $B=1T$.
In the inset the experimental configuration is shown. The XMCD
results are reported as a difference of the XAS measurements with
Right (RCP) and Left (LCP) polarizations and without any further
normalization. (b) Hysteresis loops curves between $B = -1T$ and
$B = 1T$, for 100 u.c. and 30 u.c. thick films on LAO. The curves
are normalized to unity for a better comparison of the coercitive
fields. All measurements in (a) and (b) were performed at
temperature of T=10K.}
\end{figure}

\begin{figure}
\includegraphics[width=8 cm]{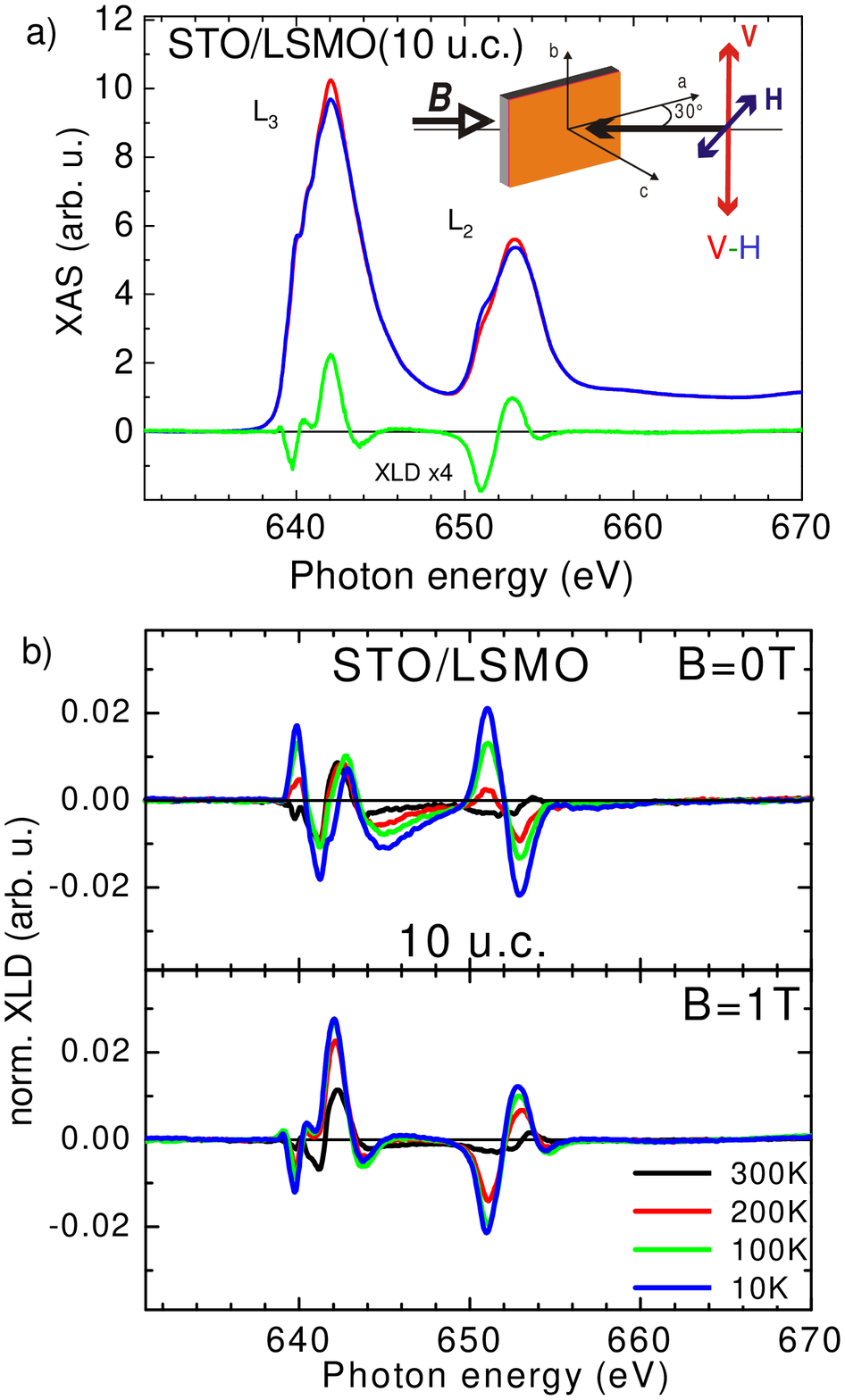}
\caption{(Color online)(a) Typical XAS and XLD (times 4) results
of 10 u.c. thick LSMO film grown on STO substrate at temperature
10K with the experimental configuration shown in the inset. XLD
spectra are reported as the difference of the XAS measurements
with Vertical (V) and Horizontal (H) polarizations, without any
further normalization. (b) Normalized XLD measurements without
(top panels) and with (bottom panels) an external magnetic field
$B=1T$ and at different temperatures ranging from 300K to 10K. The
spectra are normalized to the sum of the XAS $L_{3}$ peak height
signals.}
\end{figure}

\begin{figure}
\includegraphics[width=8 cm]{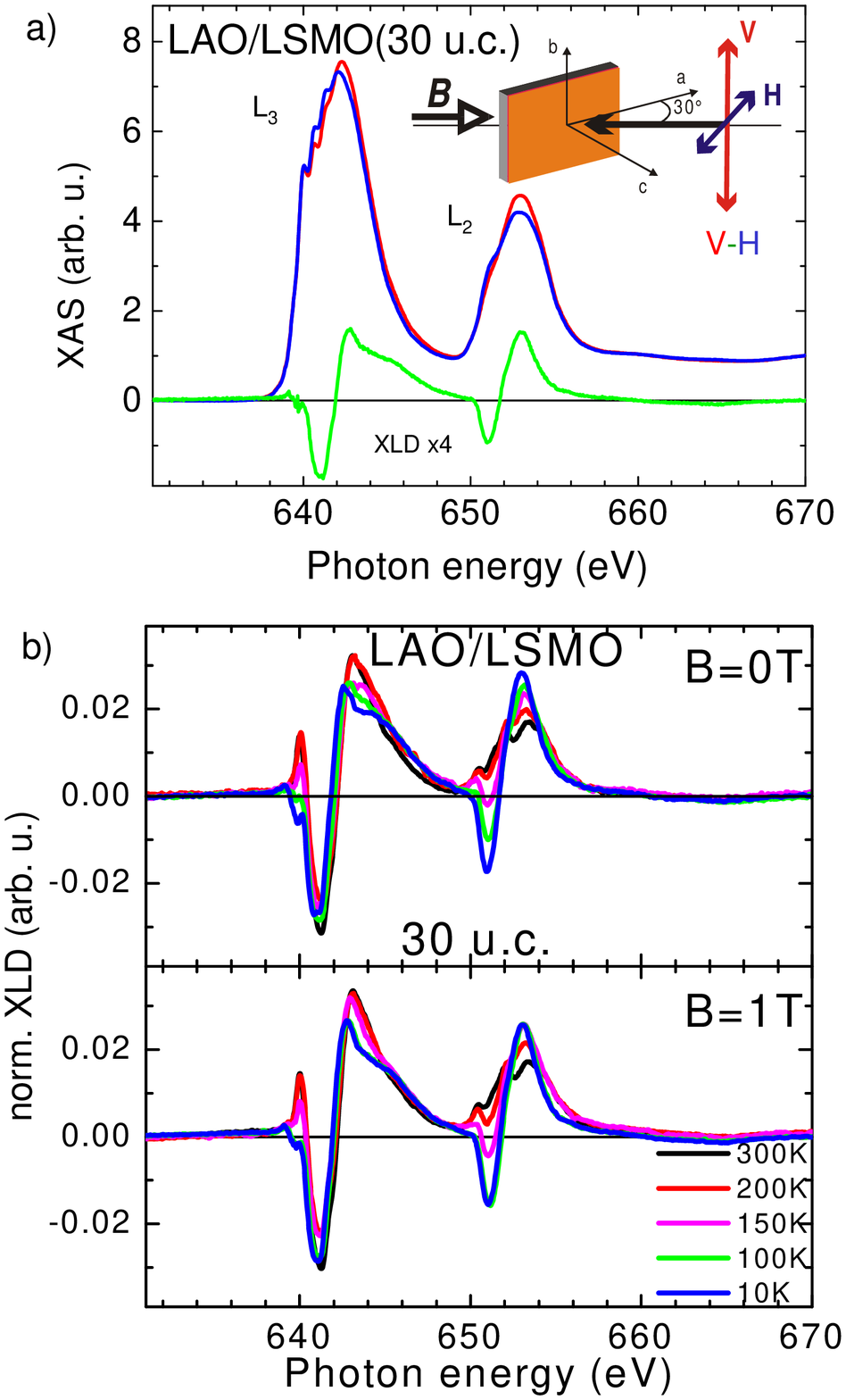}
\caption{(Color online)(a) Typical XAS and XLD (times 4) results
of 30 u.c. thick LSMO film grown on LAO substrate at temperature
10K with the experimental configuration shown in the inset. XLD
spectra are reported as the difference of the XAS measurements
with Vertical (V) and Horizontal (H) polarizations, without any
further normalization. (b) Normalized XLD measurements without
(top panels) and with (bottom panels) an external magnetic field
$B=1T$ and at different temperatures ranging from 300K to 10K. The
spectra are normalized to the sum of the XAS $L_{3}$ peak height
signals.}
\end{figure}
The case of circular dichroism is simpler: the signal is
proportional to the projection along the photon propagation
direction of the sample magnetization vector, and, at the Mn
$L_{2,3}$ edge, the effect is very strong (Fig.~2a). Thanks also
to its exceptional sensitivity (it can probe FM samples down to a
fraction of monolayer thick), XMCD can be used as an element
specific magnetometric technique, as shown by the hysteresis loops
in Fig.~2b, obtained by the maximum peak intensity of the XMCD
(about 642 eV) as a function of the applied magnetic field. In
principle, orbital and spin magnetic moments can be obtained from
the analysis of the XMCD spectra. \\On the other hand, it has been
shown theoretically and experimentally \cite{Haverkort,
vanderLaan} that XLD can have two different contributions, either
magnetic or related to the orbital occupation. If the direction of
the spin system has a component in the plane perpendicular to the
propagation direction of the x-ray beam it is possible, by
changing the linear polarization from H to V, to observe a
magnetic dichroic signal (XMLD), which can be non-zero also in the
case of AF ordering\cite{XMLD}. Furthermore, if the Mn $3d$
orbitals are anisotropically populated, in addition to the
magnetic contribution, an orbital contribution shows up in the XLD
spectra.

\section{Results}
The orbital contribution to XLD in Mn$^{3+}$ is caused by the
anisotropy in the bonding and is strictly related to the
occupation of the $e_{g}(3z^{2}-r^{2})$ or $e_{g}(x^{2}-y^{2})$
orbitals. Because the magnetic order vanishes above the magnetic
order temperature, XLD measurements performed at room temperature
are only sensitive to the preferential orbital occupation. Room
temperature XLD curves of Fig.~4b and Fig.~5b are typical of the
$e_{g}(3z^{2}-r^{2})$ preferential occupation induced by the
interface symmetry in very thin films\cite{TebanoPRL}, whatever
the sign of the mismatch between film and substrate. Furthermore,
below the magnetic transition temperature, by applying a magnetic
field parallel to the incident photon beam , it is possible to
suppress selectively the FM contribution to the XMLD spectrum thus
singling out the AF contribution. In such a geometry, if the
applied field is strong enough, magnetization in the FM system is
forced to align along the direction of the incident beam.
Therefore, the FM contribution to the XMLD is suppressed because
the spin system is orthogonal to both the V and H polarization
directions. On the other hand, spin orientation in the AF phase is
not affected by an external field, so that no major changes in the
XMLD spectra under external field are expected for the AF
phase.\\Normalized XLD spectra as a function of temperature for
LSMO films grown on STO and LAO are reported in Fig.~4b and
Fig.~5b, taken with or without the applied magnetic field $B=1$~T.
Such a field is strong enough to saturate the magnetization of the
FM phase, as demonstrated by the hysteresis loops reported in
Fig.~2b and Fig.~3b. The normalized XLD spectra for a 10 u.c.
thick LSMO film on STO is shown in Fig.~4b, with (bottom panel) or
without (top panel) applied magnetic field. As reported in Table
I, the 10 u.c. thick film on STO has a depressed $T_{\textrm{MI}}$
and becomes metallic only below 275 K, because of the proximity to
the critical thickness for the suppression of the magnetotransport
properties (about 7 u.c. for films grown on STO and
NGO)\cite{TebanoPRL}. XLD spectrum measured at 300K, above the
magnetic transition temperature, is scarcely influenced by the
application of a 1 T magnetic field, as expected. The intensity
and shape of the XLD curves changes dramatically when the film is
cooled below the magnetic ordering temperature. Large intensity
changes of the XLD spectra with temperature reveal the magnetic
dependence of the dichroic signal. Furthermore, below the ordering
temperature, the application of a 1 T field along the X ray beam
direction results in a full reversal of the XLD curves relative to
the zero field case. Such an effect is a consequence of the field
induced suppression of the FM contribution to the XLD signal which
is left with the AF contribution alone. The substantial presence
of FM phase even in a sample with reduced $T_{\textrm{MI}}$ is
confirmed by the XMCD measurements of Fig.~2a and b. The observed
changing with temperature of the XLD shape is related to the
different easy-axis orientation, i.e. local magnetic moment
preferential orientation, of the prevalent
magnetic phase at the corresponding temperature. \\
A different behavior is shown in fig.5b for the 30 u.c. thick film
on LAO. In this case, the sizeable in-plane compressive epitaxial
strain induced by the substrate strongly affects the
magneto-transport properties\cite{TebanoPRB, TebanoPRL}, thus
resulting in an insulating behaviour over the whole temperature
range. The XLD spectra at $B = 0$~T and $B = 1$~T are similar at
all temperatures indicating that the signal comes here mainly from
an AF phase. Although, the XMCD measurements of Fig.~3a and b
demonstrates the presence of a sizeable FM contribution even in
this sample, in spite of the depressed magneto-transport
properties. The domains dispersion of such FM metallic phase is
supposed to be below the percolation limit for the charge
transport because of the insulating character of the 30 u.c.  LSMO
film on LAO. The additional presence at the interface of the
$Mn^{4+}$-rich FM insulating (FMI) phase\cite{Bibes, Sidorenko, TebanoPRB} has to be also considered.\\
In order to strengthen the scenario outlined above, we have
subtracted the orbital contribution to the XLD spectra. To do this
we assumed that the orbital contribution to XLD is negligibly
sensitive to the temperature, and plotted the difference between
the XLD spectra measured below and above the magnetic ordering
temperature: $I_{XMLD} = XLD_{10K} - XLD_{300K}$, where $I_{XMLD}$
is the magnetic part of the linear dichroism signal. The
$I_{XMLD}$ spectra for LSMO samples with different thickness on
STO and on LAO are shown in Fig.~6. XMLD spectra with B=0T are
sensitive to both the AF and FM phases, while with B=1T only the
contribution of the AF phase is detected. 1 T is enough to
saturate the FM phase in the given geometry as demonstrated by the
hysteresis loops of Figs. 2b and 3b.
\begin{figure}
\includegraphics[width=8 cm]{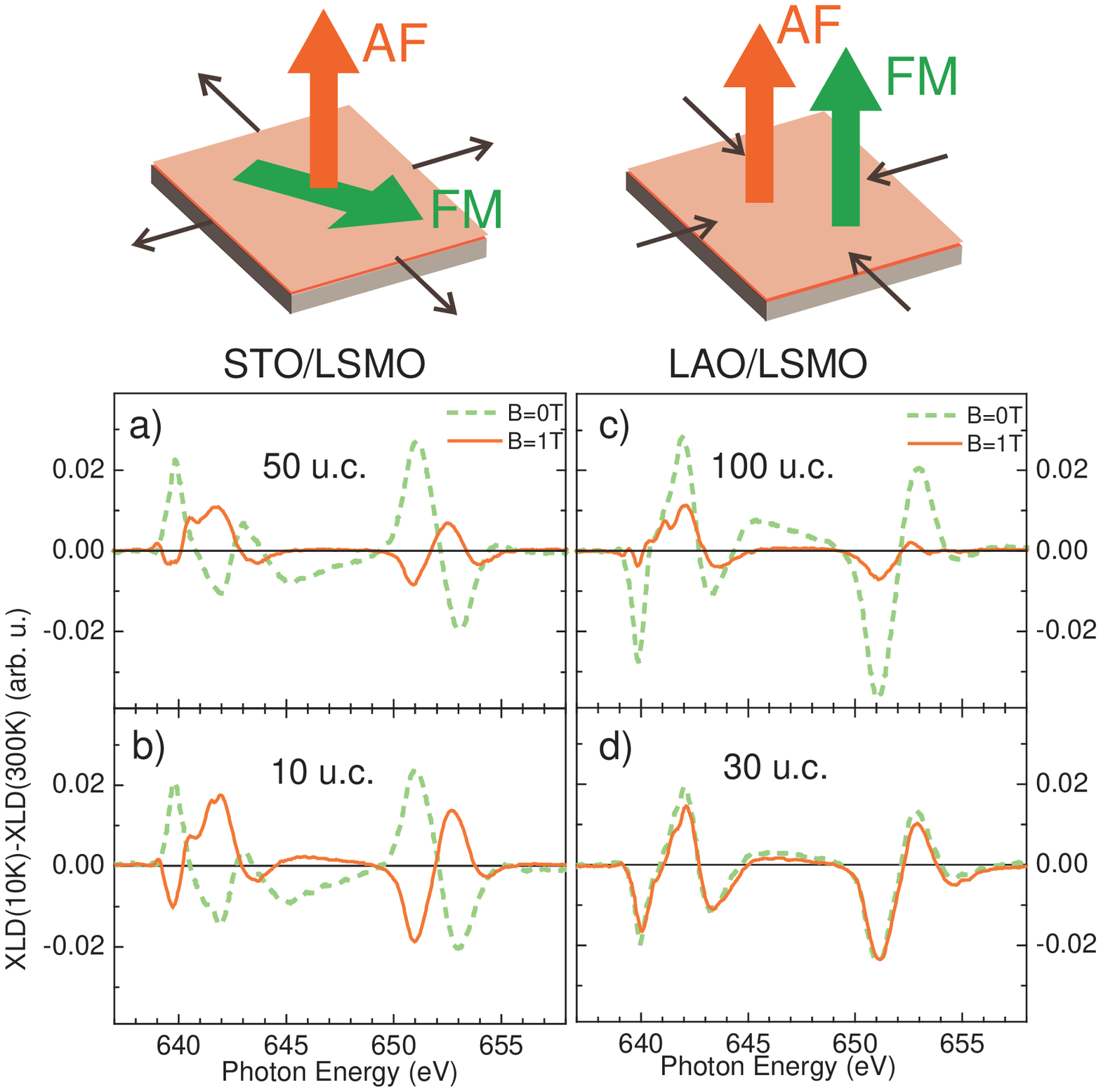}
\caption{(Color online) Difference between the XLD spectra taken
at 10K and 300K, with B=0T and B=1T, for the LSMO films: (a) 50
u.c. on STO, (b) 10 u.c. on STO,  (c)100 u.c. on LAO, (d) 30 u.c.
on LAO (the same films of Fig.3). All spectra are normalized to
the sum of the XAS $L_{3}$ peak height signals. The schematics of
the magnetization easy axes directions are reported on the top of
the figure for STO and LAO substrates, on the left and on the
right, respectively.}
\end{figure}
As reported in Table I, the 50 u.c. thick film on STO (Fig.~6a)
and the 100 u.c. thick film on LAO (Fig.~6c) are both metallic
above room temperature. The sizeable field induced suppression of
the difference spectrum in the case of the thicker film on LAO
(Fig.~6c) indicates the predominance of the FM phase. On the
contrary, when the film thickness is decreased (30 u.c. on LAO,
Fig.~6d), the $I_{XMLD}$ amplitudes at $B = 0$~T and $B = 1$~T
become comparable, in agreement with the predominance of the AF
phase. Moreover, it can be noticed that the FM and AF signals have
the same qualitative behavior, which is a clear indication that
the spin system has the same orientation in the two cases. On the
other hand, it has been reported that the magnetization easy-axis
of manganite films grown under compressive strain (LAO substrates)
is perpendicular to the substrate \cite{Nath}. Therefore, both the
FM and AF easy axes are perpendicular to the substrate ($c$-axis).
Such a finding, in agreement with the $e_{g}(3z^{2}-r^{2})$ nature
of the orbital contribution to XLD, confirms the development of
the C-type AF phase, where spins are perpendicular to the
substrate. For the two LSMO films on STO at $B = 0$~T the curves
are completely reversed with respect to those of films grown on
LAO. This fact confirms that the easy magnetization axis of the FM
phase is directed in the $ab$-plane, as already reported in
literature\cite{Nath}. On the contrary, the difference spectra at
$B = 1$~T (when the FM phase is suppressed) have the same behavior
as for the films on the LAO substrate (and on the NGO substrate
too, not shown here). As a consequence, in LSMO films grown on STO
and NGO the C-type AF phase is stabilized regardless of the small
in-plane tensile strain, which would rather be expected to favor
the A-type AF phase \cite{15, ArutaPRB, Abad, Infante}. Therefore,
in thin films grown on STO and NGO the FM easy axis lays in the
$ab$-plane whereas the AF easy axis is along the $c$-axis. The
schematic drawings of the
different easy axes directions are reported at the top of Fig.~6.\\
\begin{figure}
\includegraphics[width=8 cm]{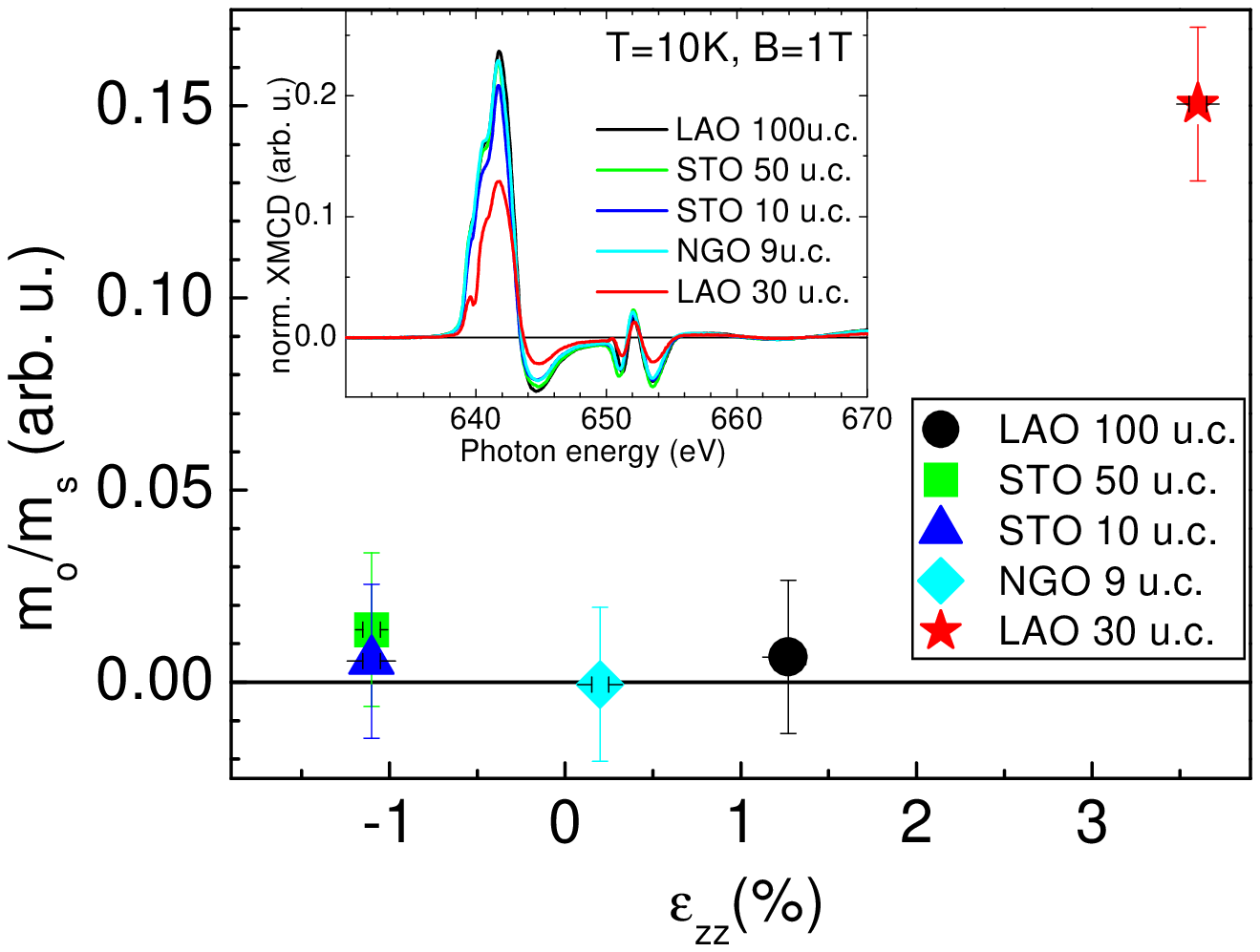}
\caption{(Color online) Orbital to spin moment ratio as a function
of the out-of-plane strain for all investigated LSMO films on STO,
NGO and LAO substrates. The corresponding XMCD spectra, normalized
to the sum of the XAS $L_{3}$ peak height signals, are reported in
the inset.}\end{figure}\\Further insight in the magnetic
properties of very thin manganite layers can be obtained from XMCD
measurements. The spin and orbital magnetic moments per atom,
$m_{s}$ and $m_{o}$, can in principle be quantified by applying
the sum rules\cite{Stohr1,Stohr2,Thole}. According to them,
$m_{s}$ and $m_{o}$ are directly related to the dichroic
difference intensities A and B (Fig.2a and 3a), which are the
$L_{3}$ and $L_{2}$ areas, respectively, of the XMCD spectra:

\begin{displaymath}
m_{s}\sim\frac{A-2B}{C}
\end{displaymath}
and
\begin{displaymath}
m_{o}\sim\frac{A+B}{C}
\end{displaymath}

where C is the XAS energy integral over the $L_{2,3}$ edges.
However, the quantitative analysis with the sum rules is
demonstrated to fail in case of Mn atoms \cite{refvanderLaan},
because of the mixing of the $L_{3}$ and $L_{2}$ core levels and
of the contribution of the magnetic dipole term. In Fig.~7 we
report the ratio of the orbital to spin moment ($m_{o}$/$m_{s}$)
as a function of the strain to qualitatively compare the
investigated samples. It is also important to underline that the A
and B values extracted from the XMCD data strongly depend on the
energy range chosen for the integration of the $L_{3}$ and $L_{2}$
edges. Therefore, the determination of the orbital and spin
moments is affected by a large uncertainty and can be carried out
with the limited purpose of a comparison among homogenous
measurements. In the inset of Fig.~7 the shape of the XMCD spectra
is shown as a whole to highlight the different behaviors of the
$L_{3}$ and $L_{2}$ edges. It follows that $m_{s}$ decreases with
the degradation of the magneto-transport properties, in agrement
with a decrease of the FM phase content, while $m_{o}$ is zero
within our experimental error in all the samples except the 30
u.c. thick film on LAO.

\section{Discussion}

Our XMLD and XMCD results can be explained in terms of the 3d
orbital occupation and the coupling among lattice distortions and
atomic moments. While XMCD technique was employed to detect the FM
spin content, the complementary XMLD technique was
used to investigate the anisotropy of the FM and AF phases.\\
By XMCD measurements, we observed a significant FM contribution
also in the insulating 30 u.c. thick LSMO film on LAO, while such
contribution was not detected by XMLD measurements. This finding
can be an indication of the in-plane orientation of the FMI phase,
thus being orthogonal to the out-of-plane-oriented FM metallic
phase in LSMO films grown on LAO. Indeed, if the FMI and FM
metallic phase are of similar amount they cancel out, as in the
case of the 30 u.c. thick film. On the contrary, in the 100 u.c.
sample the FM metallic dominates and a large difference between
the B=0 and B=1T spectra can be observed in XMLD of fig.6. Using
XMLD we found (see Fig. 6) that at the interfaces the AF C-type
phase is nucleated by the stabilization of the $3z^{2}-r^{2}$
orbital due to the break of the symmetry along the c-axis. This
also leads to preferential spin orientation out of the $ab$ plane
in the AF phase, irrespective of the strain induced by the
substrate. On the contrary in the FM regions we found a
preferential orbital occupation within the $ab$-plane for tensile
strain (STO substrate) and out of the $ab$-plane for compressive
strain (LAO). As already reported in literature\cite{Nath}, these
results can be explained in terms of the positive magnetostriction
which induces the FM easy-axis along the tensile strain direction.
We can rule out the shape anisotropy contribution in agreement
with previous reports\cite{Song}, because in our films the FM
easy-axis in-plane orientation is not strictly dependent on the
thickness of the film.
\\Moreover the evolution of $m_{o}$/$m_{s}$ measured by XMCD
reported in Fig.7 demonstrates the lattice distortion effect on
the orbital moment. It can also be explained in the framework of
coexisting  clusters  of  AF  and  FM phases, plus a minority
contribution from a FMI phase. In fact for cubic crystal field the
orbital moment is expected, from elementary considerations, to be
totally quenched \cite{Figgis}. That is what happens when FM phase
is predominant: the crystal field tetragonal distortion is much
smaller than the intrinsic width of the Mn states projected onto
the O 2p band\cite{ArutaPRB} and the orbital moment in the FM
phase remains negligible despite local distortions arising from
strain. When the thickness is reduced and the strain becomes more
and more important the AF fraction increases and the total spin
moment detected by XMCD decreases, because XMCD is insensitive to
AF moments. When the AF phase dominates (30 u.c. on LAO), a
significant elongation of the octahedra along the $z$-direction
takes place also in the FM phase, the $e_{g}(3z^{2}-r^{2})$
orbitals get preferentially occupied and the orbital moment
quenching is partially lifted \cite{Koide,Song} because of the
relevant contribution of the FMI phase. The non quenched $m_{o}$
value can also be explained in terms of the $e_{g}$ occupancy
decrease, as in the case of $Mn^{4+}$ increasing of the
interfacial FMI phase and the $e_{g}$ -band width narrowing
\cite{Koide}. The last being also related to the lattice
distortions and the more ionic character of the FMI phase. All
these effects cooperatively contribute to deviate the orbital
moment from the quenching in the thinnest LSMO film on
LAO.\\Finally we note that the AF and FM phases have rather
independent magnetic anisotropy. From one side, the non quenched
in-plane $m_{o}$ value is related to the FMI in-plane easy
magnetic axis orientation. On the other side, in the FM metallic
phase the magnetization orientation is determined by the strain
induced magnetostriction. On the contrary, in the AF phase the
preferential orbital occupation always leads to an out-of-plane
spin orientation, irrespective of the strain conditions (Fig. 6).
Therefore, in the phase separated state the exchange bias between
the AF and FM regions can be considered negligible and the spin
alignment decoupled in the case of LSMO films on STO. However, we
can guess that the FM/AF exchange-bias coupling is very small also
in the case of LAO, because the FMI phase is supposed to be
in-plane oriented, thus orthogonal to both the metallic FM and the
AF phases. In addition, despite the same spin orientation of the
AF and the metallic FM phases and the higher coercitive fields,
the presence of a relevant exchange bias should have induced an FM
hysteresis loop
shift.\\

\section{Conclusion}
We have experimentally determined the microscopic origin of
magnetic anisotropy in phase separated LSMO thin films. The
interfacial $e_{g}(3z^{2}-r^{2})$ orbital occupation favors the
the C-type AF spin ordering. Thus, the easy-axis of the AF phase
is preferentially oriented perpendicularly to the $ab$-plane for
all the substrates, whatever the sign and the strength of the
mismatch. On the contrary, in the FM phase the in-plane orbital
magnetic moment is partially unquenched when the FMI content
becomes relevant. In this case, the tetragonal distortion and the
$e_{g}$ -band width narrowing relax the quenching of the orbital
moment, giving rise to an effective spin-orbit coupling. This
demonstrates that, in the magnetic coexisting phases, the
spin-orbit to lattice coupling properties are different and
magnetic anisotropy is quite independent.

\subsection*{Acknowledgment}
Fruitful discussions with V. Iannotti and A. Galdi are acknowledged.\\

\end{document}